\documentstyle[11pt,moriond,epsfig,floatfig]{article}
\begin{document}
\initfloatingfigs
\vspace*{4cm}
\title{INCLUSIVE PARTICLE PRODUCTION AT LEP}

\author{Vladimir UVAROV}

\address{Institute for High Energy Physics, Protvino, Moscow region, Russia}

\maketitle\abstracts{
New results on inclusive production of $\Sigma^-$ and $\Lambda$(1520) 
and on proton production in quark and gluon jets are presented. These 
results are based on 2 million hadronic Z decays collected with the DELPHI 
detector at LEP. They are compared with the results of other LEP experiments 
and with models. It has been shown that the total production rates of all 
light-flavour hadrons measured so far at LEP\,1 follow phenomenological 
laws related to the spin, isospin, strangeness and mass of the particles. 
A significant proton enhancement in gluon jets is observed, indicating that 
baryon production proceeds directly from colour objects.}

\section{Introduction}

At LEP it has been shown that a large fraction of the mesons without orbital
angular momentum ($L$=0) stem from decays of scalar and tensor mesons 
($L$$\ne$0). For baryons this is not yet proven, as $L$$\ne$0 baryons typically 
have a large decay width and complicated decay modes. Hence these states are 
difficult to access experimentally in a multihadronic environment. In any case,
it is still a question of basic importance as to how far baryon production 
leads to excited baryonic states. So far the only $L$$\ne$0 baryon measured in 
$e^+e^-$ annihilation is the $\Lambda$(1520)~\cite{argus,O6,D9}.

The different colour charge of quarks and gluons leads to specific differences
in the particle production properties of the corresponding jets. Beyond the 
study of these differences~\cite{cqgj1}, which are related to the perturbative 
properties of QCD fields, the comparison of quark and gluon jets~\cite{cqgj2} 
opens up the possibility to infer properties of the non-perturbative formation 
of hadrons.

\section{Results and Discussion}

The DELPHI analysis of the inclusive production of $\Sigma^-$ and 
$\Lambda$(1520) and of the proton production in quark and gluon jets,
using 2 million hadronic Z decays recorded at LEP during 1994--95, 
has recently been published~\cite{D9,cqgj2}.

The $\Sigma^-$ is directly reconstructed as a charged track in the DELPHI 
microvertex detector and is identified by its $\Sigma^- \rightarrow {\rm n}
\pi^-$ decay leading to a kink between the $\Sigma^-$ and $\pi^-$ tracks. For 
reconstruction of the $\Lambda$(1520) resonance in the pK$^-$ mass spectrum,
tight selection criteria were required to achieve the highest possible 
purities of particle identification using the DELPHI barrel Ring Imaging 
Cherenkov detectors and the ionisation loss measurement of the Time Projection 
Chamber. The production rates per hadronic Z decay including charge conjugated 
states are measured to be:
$$ \langle \Sigma^- \rangle = 0.081 \pm 0.002 \pm 0.010 ~~~~~~~~~~ {\rm and} 
~~~~~~~~~~ \langle \Lambda(1520) \rangle = 0.029 \pm 0.005 \pm 0.005.$$
The differential distributions have been measured for both particles. 
In Ref.~\cite{D9} they are compared to the OPAL results~\cite{O7,O6} and to 
predictions of tuned~\cite{D9} Jetset 7.4 and Herwig 5.9 models. 
$\Lambda$(1520) production has been implemented in these models either by 
replacing the $\Sigma^{*0}(1385)$ by the $\Lambda$(1520) in the case of Jetset,
or by adding the $\Lambda$(1520) to the particle list in the case of Herwig. 
The model predictions have been renormalised to the observed $\Lambda$(1520) 
rate. The general shape of the $\Lambda$(1520) fragmentation function is 
reproduced well by both models. At low $x_E$, DELPHI and OPAL~\cite{O6} 
measurements agree within errors, but for $x_E > 0.3$ the DELPHI rate is about 
three times higher.

\begin{floatingfigure}{0.4\textwidth}
\centering\mbox{
\epsfig{file=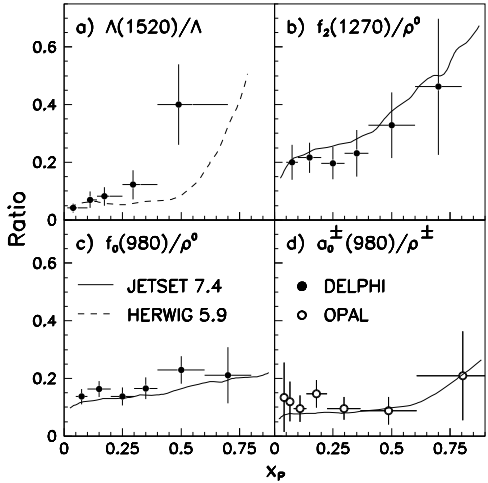,width=0.4\textwidth}}
\caption{Ratios of the differential $x_p$ distributions of $L$=1 and $L$=0
hadrons.}
\label{fig1}
\end{floatingfigure}
Fig.\,\ref{fig1}a shows the ratio of $\Lambda$(1520) to $\Lambda$ production 
as a function of the scaled momentum $x_p$. For this comparison the 
$\Lambda$ measurement~\cite{A1} is taken, as it covers a similar range in $x_p$ 
to the $\Lambda$(1520) measurement. It is seen that at small $x_p$, 
$\Lambda$(1520) production is about a factor 20 less than $\Lambda$ production.
At large $x_p$, this reduces to a factor $\sim$2.5. A similar behaviour was 
found~\cite{D8} for the ratio of tensor to vector meson production, 
$f_2(1270)/\rho^0$ (Fig.\,\ref{fig1}b). However, no increase is seen for the 
ratios of scalar to vector meson production~\cite{D8,O9}, 
$f_0(980)/\rho^0$ (Fig.\,\ref{fig1}c) and 
$a_0^{\pm}(980)/\rho^{\pm}$ (Fig.\,\ref{fig1}d).

Such a behaviour would be expected from general fragmentation dynamics due to 
the higher mass of the $\Lambda$(1520). An increase of this ratio with $x_p$
is also expected if many $\Lambda$'s stem from resonance decays. Finally 
it is interesting to note that the ratio of $\Lambda$(1520) to proton production
is identical, within errors, at low energies~\cite{argus}
and in hadronic Z decays (as calculated from Refs.~\cite{D9,Dadd}).

To check for a possible spin alignment of the $\Lambda$(1520), the 
distribution of the cosine of the kaon angle in the $\Lambda$(1520) rest system 
with respect to the $\Lambda$(1520) direction has been measured~\cite{D9} for 
$x_p > 0.07$ and fitted with the expected form, yielding the spin density 
matrix element value
$$\rho_{{+}{1 \over 2}{+}{1 \over 2}} + \rho_{{-}{1 \over 2}{-}{1 
\over 2}} = 0.4 \pm 0.2.$$
Thus no significant $\Lambda$(1520) spin alignment is observed. 

The total production rates of light-flavour hadrons in hadronic Z decays were 
measured at least for one state of an isomultiplet in ALEPH~\cite{A1,Aall}, 
DELPHI~\cite{D9,D8,Dadd,Dall}, L3~\cite{Lall} and OPAL~\cite{O6,O7,O9,Oall}.
It has been shown that the total production rates of vector, tensor and 
scalar mesons~\cite{Panic99} and of baryons~\cite{D9} follow phenomenological 
laws related to the spin ($J$), isospin ($I$), strangeness ($S$) and mass 
($M$) of the particles. The main idea of this analysis was to plot the 
baryon and meson production rates in a different way. For baryons we analyse 
{\it the sum of the production rates} of all states of an isomultiplet as a 
function of $M^2$. In the case that not all states of an isomultiplet are 
measured at LEP, equal production rates for the other states is assumed. For 
mesons we analyse {\it the production rates per spin and isospin projection} 
as a function of $M$. These rates were obtained by averaging the rates of 
particles belonging to the same isomultiplet, excluding charge conjugated 
states and divided by a spin factor (2$J$+1).

\begin{figure}
\centering\mbox{
\epsfig{file=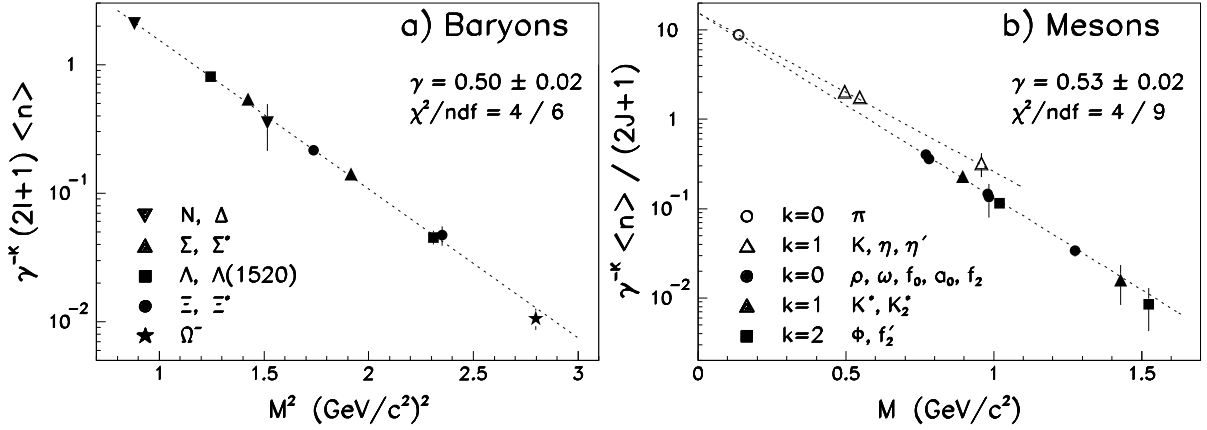,width=\textwidth}}
\caption{The weighted production rates of baryons as a function of $M^2$ and 
of mesons as a function of $M$.}
\label{fig2}
\end{figure}

The total production rates of all light-flavour hadrons measured so far 
at LEP\,1 are well fitted by the formulas 
$$(2I+1)\,{\langle n \rangle} \,\equiv\,
{\Sigma_i}\,{\Sigma_j}\,{\langle n \rangle}_{ij}
\,=\, A\,\gamma^k\,\exp{(-b\,M^2)} ~~~{\rm and}~~~
{\langle n \rangle}\,/\,(2J+1) \,\equiv\,
{\langle n \rangle}_{ij} \,=\, A\,\gamma^k\,\exp{(-b\,M)}^{~}$$
for baryons and mesons respectively, where ${\langle n \rangle}_{ij}$ is the 
production rate per spin and isospin projection and $k$ is the number of $s$
and $\bar{s}$ quarks in the hadron. The values of the fitted 
\begin{table}[hp]
\small
\begin{center} 
\begin{tabular}{|ll|c|c|c|c|}
\hline
Particles & & $A$ & $b$ & $\gamma$ & $\chi^2 / ndf$ \\ 
\hline
Baryons & $(B)$  
& 22.1$\pm$1.6 & 2.66$\pm$0.08 & 0.50$\pm$0.02 & 4.1 / 6 \\
Vectors, Tensors, Scalars & $(V,T,S)$ 
& 18.6$\pm$4.3 & 4.95$\pm$0.26 & 0.53$\pm$0.02 & 2.2 / 6 \\
Pseudoscalars & $(P)$ 
& 15.1$\pm$1.2 & 3.87$\pm$0.57 & 0.49$\pm$0.10 & 0.5 / 1 \\
\hline
\end{tabular}
\label{tab1}
\end{center}
\normalsize
\end{table}
parameters are given~\footnote{$~b$ in (GeV/$c^2$)$^{-2}$ for baryons and in 
(GeV/$c^2$)$^{-1}$ for mesons.} in the Table. The factor $\gamma$ is the same
for all hadrons: $\gamma_{\,average} = 0.51 \pm 0.02$.
If the production rates are weighted by $\gamma^{-k}$, a universal mass
dependence is observed for all baryons (Fig.\,\ref{fig2}a). But for
mesons there are two dependences: one for pseudoscalar mesons and another   
for vector, tensor and scalar mesons (Fig.\,\ref{fig2}b).
The slopes $b$ are different for mesons with net spin 0 and 1, but they are 
the same for baryons with net spin $1 \over 2$ and $3 \over 2$. The slopes $b$ 
do not depend on the value and orientation of the {\it orbital angular 
momentum} $L$ of the quarks for mesons with net spin 1 and for baryons with 
net spin $1 \over 2$.

Using the values of the parameter $A$, the hadron ratios can be extrapolated 
to $M = 0$:
$$\rho^+ /\,3\,\pi^+ ~=~ A_{\,V,T,S}\,/\,A_{P} ~=~ 1.2\,\pm\,0.3
~~~~~{\rm and}~~~~~
\pi^+ /\,p ~=~ 4\cdot A_{M}\,/\,A_{B} ~=~ 2.8\,\pm\,0.3,$$
where $A_M = 15.4 \pm 1.2$ is the weighted average of $A_{\,V,T,S}$ and $A_P$.
These ratios agree with predictions of the quark combinatorics 
model~\cite{Anisovich},~ 
$\rho^+ : \pi^+ = 3 : 1$ and $\pi^+ : p \,:\, \bar{p} = 3 : 1 : 1$.

The inclusive distributions of $\pi^+$, $K^+$ and $p$ in hadronic Z decays
have been measured in quark and gluon jets~\cite{cqgj2}. Three-jet events were 
clustered using the Durham algorithm with a jet resolution parameter 
$y_{cut}=0.015$. To obtain samples of quark and gluon jets with similar 
kinematics, only the low energy jets for Y events and all jets for Mercedes 
events were used. Gluon jets were selected in $b\bar{b}g$ events by 
anti-tagging the $b$ quarks using an impact parameter technique. 

The point of intersection of the $\xi_p = -\ln{x_p}$ (with 
$x_p = p_{particle}\,/\,p_{jet}$) distributions 
(see Fig.~9 in Ref.~\cite{cqgj2}) of quark and gluon jets 
is approximately the same for pions and kaons, ${\xi_p}^{(s)} \sim 1.73$. For 
protons the crossing point between the quark and gluon distributions is 
shifted to higher momentum at ${\xi_p}^{(s)} \sim 0.74$. Thus proton 
production is enhanced in gluon jets, but preferentially at high momenta. 
A surplus of baryon production in gluon jets with the observed kinematical 
properties can be qualitatively understood if baryons are directly produced 
from coloured partons or equivalently from a colour string 
(see Ref.~\cite{cqgj2} for more details).

\section{Summary}

DELPHI has recently measured total $\Sigma^-$ and $\Lambda$(1520) production 
rates per hadronic Z decay including charge conjugated states to be
0.081\,$\pm$\,0.010 and 0.029\,$\pm$\,0.007. The differential distributions 
for both particles are well described by the tuned Jetset 7.4 and Herwig 5.9 
Monte Carlo models. No significant $\Lambda(1520)$ spin alignment was observed. 

The ratio of $\Lambda$(1520) to $\Lambda$ production increases with 
increasing scaled momentum $x_p$. A similar behaviour was found
for the ratio of tensor to vector meson production, $f_2(1270)/\rho^0$. 
However, no increase was seen for the ratios of scalar to vector meson 
production, $f_0(980)/\rho^0$ and $a_0^{\pm}(980)/\rho^{\pm}$.

The total production rates of all light-flavour hadronic 
states measured so far at LEP\,1 follow phenomenological laws related to 
the spin, isospin, strangeness and mass of the particles. The hadron ratios,\, 
$\rho : \pi$ and $\pi : p$\,, extrapolated to $M = 0$ using these laws agree 
with predictions of the quark combinatorics model. 

The production spectra of the identified particles have been measured in quark 
and gluon jets. A significant proton enhancement in gluon jets is observed, 
consistent with baryon production proceeding directly from coloured objects.

\end{document}